\documentstyle[12pt]{article}
\newcommand{\pr}{\hspace{\parindent}}
\newcommand{\beq}{\begin{equation}}
\newcommand{\eeq}{\end{equation}}
\newcommand{\bea}{\begin{eqnarray}}
\newcommand{\eea}{\end{eqnarray}}
\newcommand{\bsub}{\begin{subequations}}
\newcommand{\esub}{\end{subequations}}

\def\simlt{\rlap{\lower 3.5 pt \hbox{$\mathchar \sim$}} \raise 1pt \hbox {$<$}}
\def\simgt{\rlap{\lower 3.5 pt \hbox{$\mathchar \sim$}} \raise 1pt \hbox {$>$}}
\def\Im{\mathop{{\cal I}\mskip-4.5mu \lower.1ex \hbox{\it m}}}
\def\Re{\mathop{{\cal R}\mskip-4mu \lower.1ex \hbox{\it e}}}

\def\etal{{\it et al}.\ }
\def\ibid{{\it ibid}.\ }

\def\to{\rightarrow}
\def\epem{\ifmmode{ e^{+}e^-} \else{$ e^{+}e^- $ } \fi}
\def\ttbar{\ifmmode{t\bar{t}} \else{$t\bar{t}$} \fi}
\def\eetoentb{e^- e^+ \rightarrow e^- \bar{\nu}_e t \bar{b} }
\def\eetowtb{ e^- e^+ \rightarrow W^- t \bar{b} }
\def\aetontb{ \gamma e^+ \rightarrow \bar{\nu}_e t \bar{b} }

\def\gev{{\rm\,GeV}}

\renewcommand{\theequation}{\arabic{equation}}

\makeatletter
\newtoks\@stequation

\def\subequations{\refstepcounter{equation}%
  \edef\@savedequation{\the\c@equation}%
  \@stequation=\expandafter{\theequation}%   %only want \theequation
  \edef\@savedtheequation{\the\@stequation}% %expanded once
  \edef\oldtheequation{\theequation}%
  \setcounter{equation}{0}%
  \def\theequation{\oldtheequation\alph{equation}}}

\def\endsubequations{%
  \ifnum\c@equation < 2 \@warning{Only \the\c@equation\space subequation
    used in equation \@savedequation}\fi
  \setcounter{equation}{\@savedequation}%
  \@stequation=\expandafter{\@savedtheequation}%
  \edef\theequation{\the\@stequation}%
  \global\@ignoretrue}

\def\eqnarray{\stepcounter{equation}\let\@currentlabel\theequation
\global\@eqnswtrue\m@th
\global\@eqcnt\z@\tabskip\@centering\let\\\@eqncr
$$\halign to\displaywidth\bgroup\@eqnsel\hskip\@centering
     $\displaystyle\tabskip\z@{##}$&\global\@eqcnt\@ne
      \hfil$\;{##}\;$\hfil
     &\global\@eqcnt\tw@ $\displaystyle\tabskip\z@{##}$\hfil
   \tabskip\@centering&\llap{##}\tabskip\z@\cr}

\makeatother

\begin{document}

\vspace*{-10mm}

\baselineskip12pt
\begin{flushright}
\begin{tabular}{l}
{\bf hep-ph/9401295}\\
{\bf KEK-TH-384 }\\
{\bf DTP 94/06 }\\
{\bf KEK preprint 93--183 }\\
January 1994
\end{tabular}
\end{flushright}

\baselineskip18pt
\vspace{8mm}
\begin{center}
{\Large \bf Single Top Production at LEP~200 } \\
\vglue 8mm
{\bf K.~Hagiwara,~~~M.~Tanaka } \\
\vglue 1mm
{\it Theory Group, KEK, Tsukuba, Ibaraki 305, Japan }\\
\vglue 4mm
{and} \\
\vglue 4mm
{\bf T.~Stelzer }\\
\vglue 1mm
{\it Department of Physics, Durham University, DH1 3LE, England }\\
\vglue 20mm
{\bf ABSTRACT} \\
\vglue 10mm
\begin{minipage}{14cm}
\baselineskip18pt
We present exact tree level cross sections for the single top
production process $\eetoentb$ at LEP~200.  The results reproduce
roughly those obtained earlier by using the equivalent real photon
approximation and we confirm the observation that detecting a top
heavier than half the c.m.~energy is not feasible at LEP~200.  The
calculation has been performed by a new automatic Feynman amplitude
generator MadGraph which produces HELAS code for the helicity
amplitudes.
\end{minipage}
\end{center}
\vfill
\newpage
The search for the top quark is a primary target of present and future
collider experiments.
For a top quark lighter than half the center of mass (c.m.) energy
$\sqrt{s}$, top production at \epem colliders will be dominated by
the pair production process $e^+e^- \rightarrow t\bar{t}$.
This is, however, unlikely to be the case at LEP~200 if the
CDF bound\cite{CDF}
on the top mass of $m_t \simgt 120~\gev$ is valid.
We can expect copious production of \ttbar pairs at LEP~200
only if the top decays mainly into exotic modes which would invalidate the
above bounds from hadron collider experiments.

The standard top quark can still be produced singly at LEP~200
via the higher order processes
\beq
\label{eetowtb}
\eetowtb ,
\eeq
if $m_t < \sqrt{s}-m_W-m_b$, and
\beq
\label{eetoentb}
\eetoentb ,
\eeq
if $m_t < \sqrt{s}-m_b$.
Recently it has been claimed that single top quark production
can be discovered at LEP~200 up to $m_t \sim 165~\gev$\cite{raidal}.
Another work states that the cross section is much too small
for LEP~200 to produce top quarks singly with its design
luminosity\cite{panella}.
The former calculation introduces an artificial cutoff to the single
top quark production cross section which is singular in the massless
electron limit.
The latter calculation makes use of the equivalent real photon
approximation (EPA) to estimate the cross section from that of the subprocess
\beq
\label{aetontb}
\aetontb .
\eeq

In order to resolve this conflict,
we present in this report the exact tree level cross section for the single
top quark production processes (\ref{eetowtb}) and (\ref{eetoentb}),
and compare the latter cross section with that obtained by using the EPA.
We find that the naive EPA as employed in ref.\cite{panella}
{overestimates} the cross section by about 30\% and that
one should expect even smaller cross sections than those estimated there.
Therefore, we essentially confirm the observation of
Panella~\etal\cite{panella} that the single top production rate is
too small to be interesting at LEP~200 with its planned luminosity of
0.5~fb$^{-1}/$year.

Our calculation has been performed by a recently completed automatic
Feynman amplitude generator MadGraph\cite{madgraph}.
Given the initial and final state particles for a Standard Model process,
MadGraph automatically generates a postscript file of the Feynman graphs
and a FORTRAN program of the helicity amplitudes that makes use of the
HELAS subroutines\cite{helas}.
We show in Fig.~1 the Feynman graphs of process (\ref{eetoentb})
as generated by MadGraph.
Differential cross sections can be easily obtained by integrating the
squared matrix elements over the relevant phase space range.

It is not so trivial to obtain the total cross section of the process,
because the matrix element becomes singular at high energies
when the exchanged virtual photon in graphs 3, 5, 7 and 9 of Fig.~1
becomes nearly on-shell.
The HELAS subroutines\cite{helas} numerically calculate the helicity
amplitudes accurately in the singular region by modifying the $ee\gamma$
currents in the relevant diagrams and by choosing an appropriate
parametrization of the phase space volume\cite{hagiwara} such that
no subtle cancellation occurs in the numerical program.

The results are shown in Fig.~2.  Solid lines show the total cross
sections for process (\ref{eetoentb}) and dashed lines show those
for process (\ref{eetowtb}).
In this calculation, we neglect the width of the top quark, and hence
the cross sections are given only down to $m_t =\sqrt{s}/2+2~\gev$.
Near the \ttbar pair production threshold, one should examine the
$\epem \to b\bar{b} W^+W^-$ amplitudes carefully in the presence
of the large non-perturbative QCD corrections\cite{tt}.
In the helicity amplitudes for the process $\eetoentb$,
the so-called {\it annihilation} graphs 11--20 of Fig.~1 are
a gauge invariant set.  They contribute negligibly to the total
cross section below the $\eetowtb$ threshold, while they give
just a fraction of the $\eetowtb$ cross section above the threshold.
The solid lines of Fig.~2 are obtained by consistently neglecting
the contributions from the annihilation diagrams.

It is clear from Fig.~2 that the single top quark production cross section
below the $\eetowtb$ threshold is never greater than the 0.1~fb level
at any conceivable LEP~200 energy, and that it remains below the level
of a few times 0.1~fb even above the $\eetowtb$ threshold.
These observations confirm those of ref.\cite{panella} qualitatively,
and disagree with ref.\cite{raidal}.
For the convenience of future comparison, we give some representative values
of the total cross sections in Table~1 ($\sqrt{s}=190~\gev$) and
in Table~2 ($\sqrt{s}=200~\gev$).
The SM parameters are chosen as
$m_b = 5\gev$,
$m_W = 80\gev$,
$m_Z = 91\gev$,
$e^2/4\pi = \sin^2\theta_W g^2/4\pi = 1/128$,
$\sin^2\theta_W =0.23$,
and all the widths have been set to zero.
The total cross section for the process $\eetoentb$ is then multiplied by
an overall factor of $128/137$, since the cross section is dominated by
the region of the phase space where the invariant mass of the virtual photon
that couples to the external electron current is much smaller than one
GeV \cite{hagiwara}.
Numerical errors associated with the Monte Carlo integration over the
phase space volume are estimated to be less than 1\% \cite{bases}.

A closer comparison of our exact tree level results with the results using
the equivalent real photon approximation (EPA) in ref.\cite{panella}, reveals
that our exact cross section is consistently smaller than theirs by about
30\%.  To ensure that this is due to the approximation and not an error,
we also calculate the cross section using the EPA.

Shown in Fig.~3 is the total cross section for the process
$\aetontb$ as a function of the $\gamma e$ c.m.~energy
$\sqrt{s_{\gamma e}}$.
The curves are calculated by using the MadGraph/HELAS system
with the same SM parameters.
The subprocess cross sections agree rather well with those reported
in ref.\cite{panella}.  The validity of our calculation is further
varified by comparing our result for a light top $m_t<m_W$, with the
earlier result of ref.\cite{katuya}.

Finally, in Fig.~4 we compare three estimates of the total cross section
for the process $\eetoentb$.
The solid lines are obtained by our exact tree level calculation.
The long dashed lines are obtained by using
the naive equivalent real photon distribution
\beq
\label{naiveepa}
	D_{\gamma/e}(z,s)_{\rm naive \: EPA} = \frac{\alpha}{2\pi}
	\frac{1+(1-z)^2}{z}\log\frac{s}{4m_e^2}
\eeq
as adopted by Panella \etal\cite{panella}.  And the short dashed lines are
obtained by using the improved equivalent real photon ($\overline{\rm EPA}$)
distribution
\beq
\label{epabar}
	D_{\gamma/e}(z,Q^2)_{\overline{\rm EPA}} = \frac{\alpha}{2\pi}
	\{\frac{1+(1-z)^2}{z}[\log\frac{Q^2 (1-z)}{m_e^2 z^2}-1]
	\: +\frac{z}{2}\}
\eeq
as proposed in ref.\cite{hagiwara}.
Here the $\overline{\rm EPA}$ flux of the photon is determined by respecting
the exact lower kinematical limit of the virtual photon mass squared
$t_{\rm min}=m_e^2 z^2/(1-z)^2$,
whereas for the maximal virtuality $Q^2$ consistent
with the real photon approximation, we take the typical virtuality
scale of the subprocess $\aetontb$,
\beq
\label{scale}
	Q^2 = m_b^2 -(p_{\gamma}-p_b)^2.
\eeq
The last term in the $\overline{\rm EPA}$ distribution (\ref{epabar})
without the logarithmic enhancement gives a small, but universal,
contribution from the electron helicity flip amplitudes.
Fig.~4 clearly shows that the naive EPA overestimates the cross section
mainly because it fails to take account of the relatively small
effective scale of the subprocess $Q^2$ (\ref{scale}).
The improved EPA underestimates the cross section slightly because
the exact cross section does not disappear when the virtual mass
of the external electron current exceeds the scale (\ref{scale}).
This is consistent with the $\overline{\rm EPA}$ result as reported
in ref.\cite{hagiwara} for the process $\epem \to \epem Z$.

The exact tree level cross sections of Fig.~4 (solid lines) are obtained
by using all the diagrams of Fig.~1.  One can observe effects of
the annihilation diagrams as small enhancements of the cross section
near the $\eetowtb$ threshold over the $\overline{\rm EPA}$ estimates.

We conclude that our exact calculation of the single top production
cross section is valid and that detecting a top heavier than half
the c.m.~energy is not feasible at LEP~200.

%%%%%%%%%%%%%%%%%%%%%%%%%%%%%%%%%%%%%%%%%%%%%%%%%%%%%%%%%%%%%%%%%%%%%%%%%%%%%%
\section*{\normalsize \bf Acknowledgements }
We wish to thank M.~Sasaki for calling our attention to the
discrepancy between refs.\cite{raidal} and \cite{panella}.
We would also like to thank J.~Kanzaki and I.~Watanabe for
helpful comments.

%%\input{txt}
%%\input{ref}
%%%%%%%%%%%%%%%%%%%%%%%%%%%%%%%%%%%%%%%%%%%%%%%%%%%%%%%%%%%%%%%%%%%%%%

%\end{list}
\vfill
%%\input{tab}
%%%%%%%%%%%%%%%%%%%%%%%%%%%%%%%%%%%%%%%%%%%%%%%%%%%%%%%%%%%%%%%%%%%%%%%%%%%%
\newpage
\section*{ \large Tables }
%%%%%%%%%%%%%%%%%%%%%%%%%%%%%%%%%%%%%%%%%%%%%%%%%%%%%%%%%%%%%%%%%%%%%%%%%%%%
\subsection*{ \normalsize Table 1:
Exact tree level cross sections for the process $\eetowtb$
and the process $\eetoentb$ at $\sqrt{s} = 190~\gev$ for several top masses.
The latter cross sections are obtained by neglecting contributions from the
{\it annihilation} diagrams 11--20 of Fig.~1.
The SM parameters are chosen as follows:
$m_b = 5\gev$,
$m_W = 80\gev$,
$m_Z = 91\gev$,
$e^2/4\pi = \sin^2\theta_W g^2/4\pi = 1/128$,
and $\sin^2\theta_W =0.23$.
All the widths have been set to zero.
The latter cross sections are obtained by multiplying the total cross sections
by an overall factor $128/137$\cite{hagiwara}.
}
\def\none{--------}
\begin{center}
\begin{tabular}{|c|c|l|}
\hline
	$m_t$ &
        {$\sigma(\eetowtb)$} &
	{$\sigma(\eetoentb)$} \\
\hline
	100~\gev & 0.0050~fb & \hspace{1em}0.046~fb \\
\hline
	110~\gev & {\none} & \hspace{1em}0.027~fb \\
\hline
	120~\gev & {\none}  & \hspace{1em}0.016~fb \\
\hline
	130~\gev & {\none}  & \hspace{1em}0.0091~fb \\
\hline
	140~\gev & {\none}  & \hspace{1em}0.0044~fb \\
\hline
	150~\gev & {\none}  & \hspace{1em}0.0018~fb \\
\hline
	160~\gev & {\none}  & \hspace{1em}0.00048~fb \\
\hline
	170~\gev & {\none} & \hspace{1em}0.000061~fb \\
%\hline
%	180~\gev & {\none} & \hspace{1em}0.00000054~fb \\
\hline
\end{tabular}
\end{center}
\subsection*{ \normalsize Table 2:
Same as Table~1 but for the \epem c.m. energy $\sqrt{s} = 200~\gev$.
}
\def\none{--------}
\begin{center}
\begin{tabular}{|c|c|l|}
\hline
	$m_t$ & $\sigma(\eetowtb)$ & $\sigma(\eetoentb)$ \\
%\hline
%	100~\gev & \none & \hspace{3em}\none \\
\hline
	110~\gev & 0.0022~fb & \hspace{2em}0.040~fb \\
\hline
	120~\gev & \none  & \hspace{2em}0.025~fb \\
\hline
	130~\gev & \none  & \hspace{2em}0.015~fb \\
\hline
	140~\gev & \none  & \hspace{2em}0.0085~fb \\
\hline
	150~\gev & \none  & \hspace{2em}0.0041~fb \\
\hline
	160~\gev & \none  & \hspace{2em}0.0016~fb \\
\hline
	170~\gev & \none  & \hspace{2em}0.00044~fb \\
\hline
	180~\gev & \none  & \hspace{2em}0.000056~fb \\
\hline
\end{tabular}
\end{center}
%%%%%%%%%%%%%%%%%%%%%%%%%%%%%%%%%%%%%%%%%%%%%%%%%%%%%%%%%%%%%%%%%%%%%%%%%%%
\vfill

%%%\input{fig}
%%%%%%%%%%%%%%%%%%%%%%%%%%%%%%%%%%%%%%%%%%%%%%%%%%%%%%%%%%%%%%%%%%%%%%%%%%%%
\newpage
\noindent
\section*{ \large Figures }
\renewcommand{\labelenumi}{Fig.~\arabic{enumi}}
\begin{enumerate}
%%%%%%%%%%%%%%%%%%%%%%%%%%%%%%%%%%%%%%%%%%%%%%%%%%%%%%%%%%%%%%%%%%%%%%%%%%%%
\item	%Fig.~1
Feynman diagrams for the process $\eetoentb$ as generated by the automatic
Feynman amplitude generator MadGraph\cite{madgraph}.

\item	%Fig.~2
Exact tree level cross sections for the process $\eetoentb$ (solid lines)
and the process $\eetowtb$ (dashed lines) plotted against the top mass
$m_t$ at five c.m.~energies $\sqrt{s} =$180, 190, 200, 210, and 220~GeV.
The solid lines are obtained by neglecting contributions from the
{\it annihilation} diagrams 11--20 of Fig.~1.
The SM parameters are chosen as follows:
$m_b = 5\gev$,
$m_W = 80\gev$,
$m_Z = 91\gev$,
$e^2/4\pi = \sin^2\theta_W g^2/4\pi = 1/128$,
and $\sin^2\theta_W =0.23$.
All the widths have been set to zero.
The solid lines are obtained by multiplying the total cross sections
by an overall factor $128/137$\cite{hagiwara}.

\item	%Fig.~3
Exact tree level cross sections for the process $\aetontb$
plotted against the $\gamma e$ c.m.~energy $\sqrt{s_{\gamma e}}$
for six typical top mass $m_t =$100, 120, 140, 160, 180, and 200~GeV.
The parameters are the same as those used in Fig.~2.

\item	%Fig.~4
Exact tree level cross sections for the process $\eetoentb$ (solid lines)
as compared with the cross sections obtained by the naive EPA (equivalent
real photon approximation) of ref.\cite{panella} (long dashed lines) and
those by the improved EPA of ref.\cite{hagiwara} (short dashed lines).
Here all the diagrams of Fig.~1 have been included in the exact cross sections.
The parameters are the same as those used in Fig.~2.

%%%%%%%%%%%%%%%%%%%%%%%%%%%%%%%%%%%%%%%%%%%%%%%%%%%%%%%%%%%%%%%%%%%%%%%%%%%%%
\end{enumerate}
\vfill
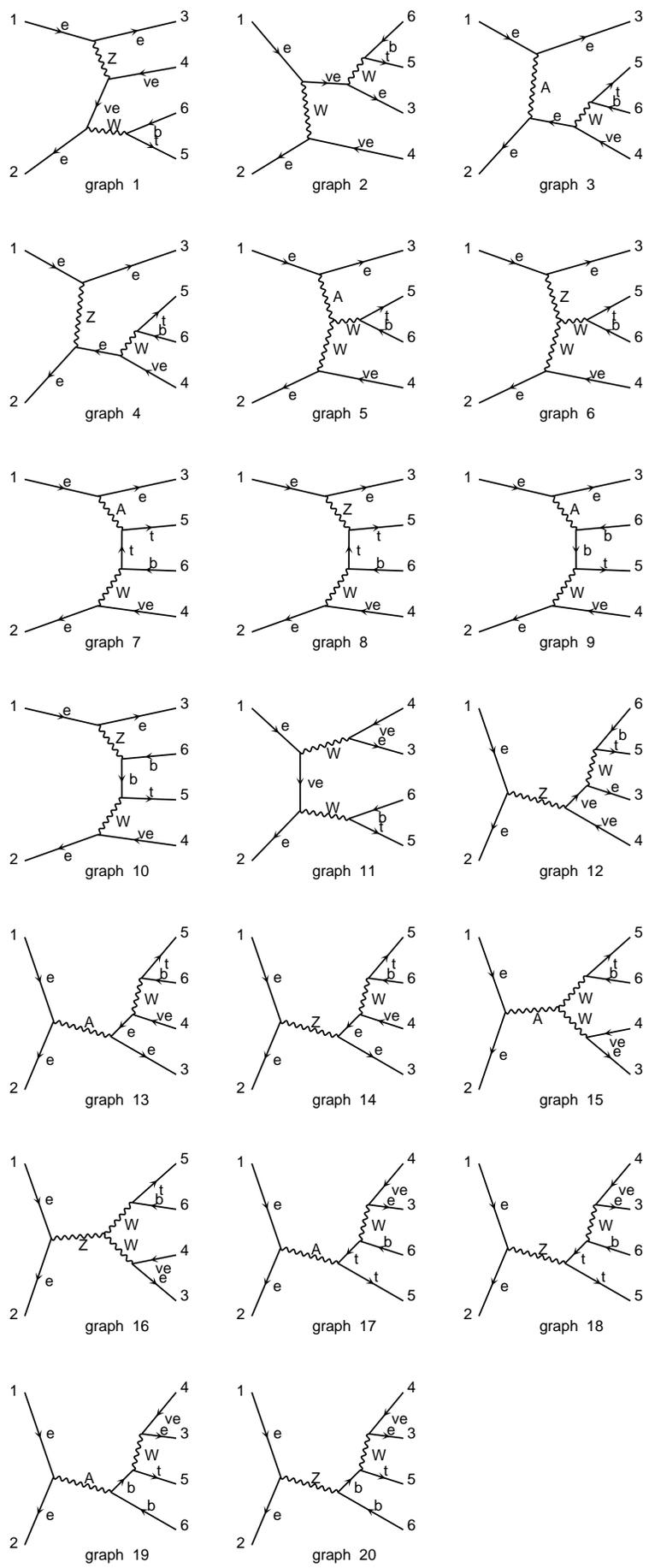
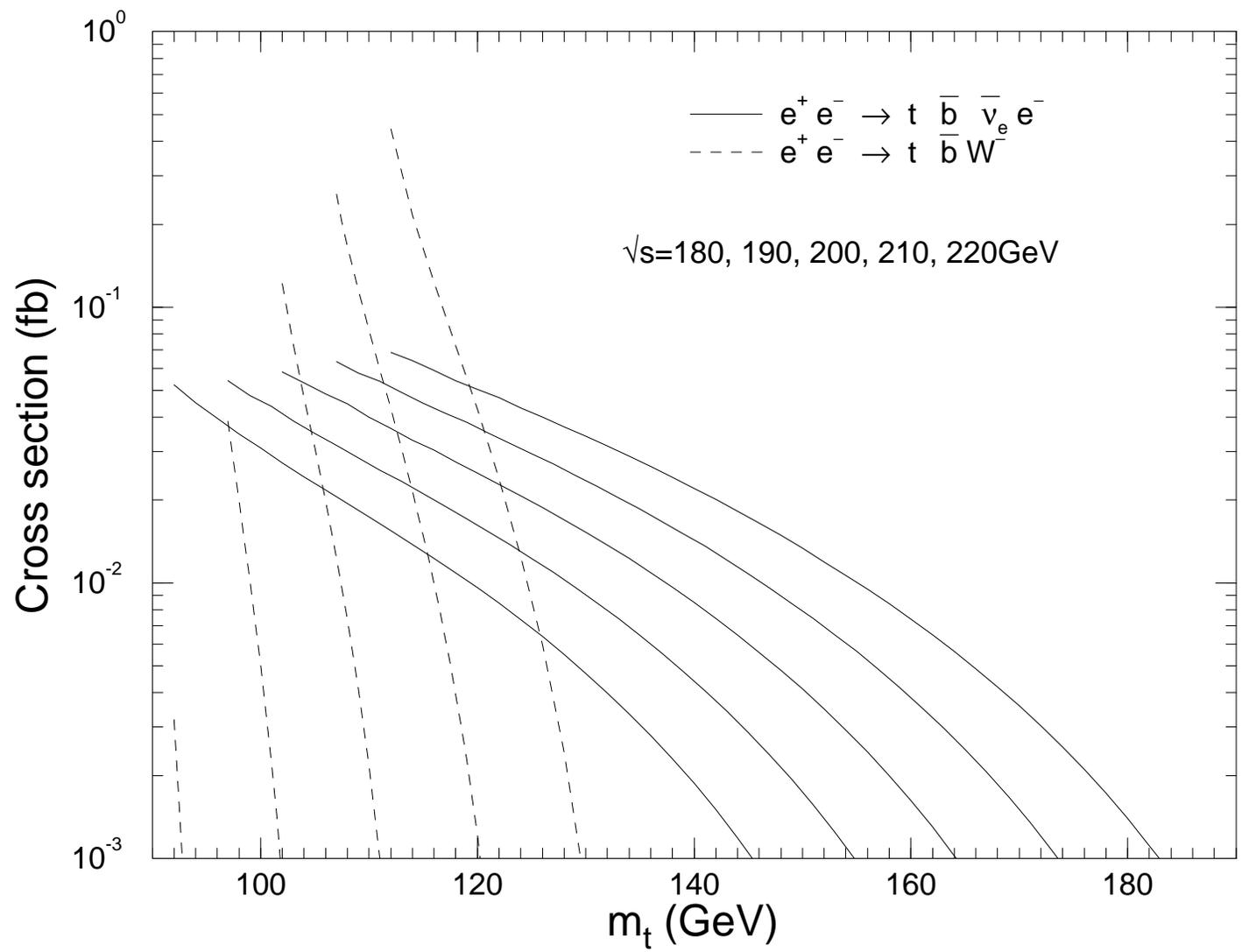
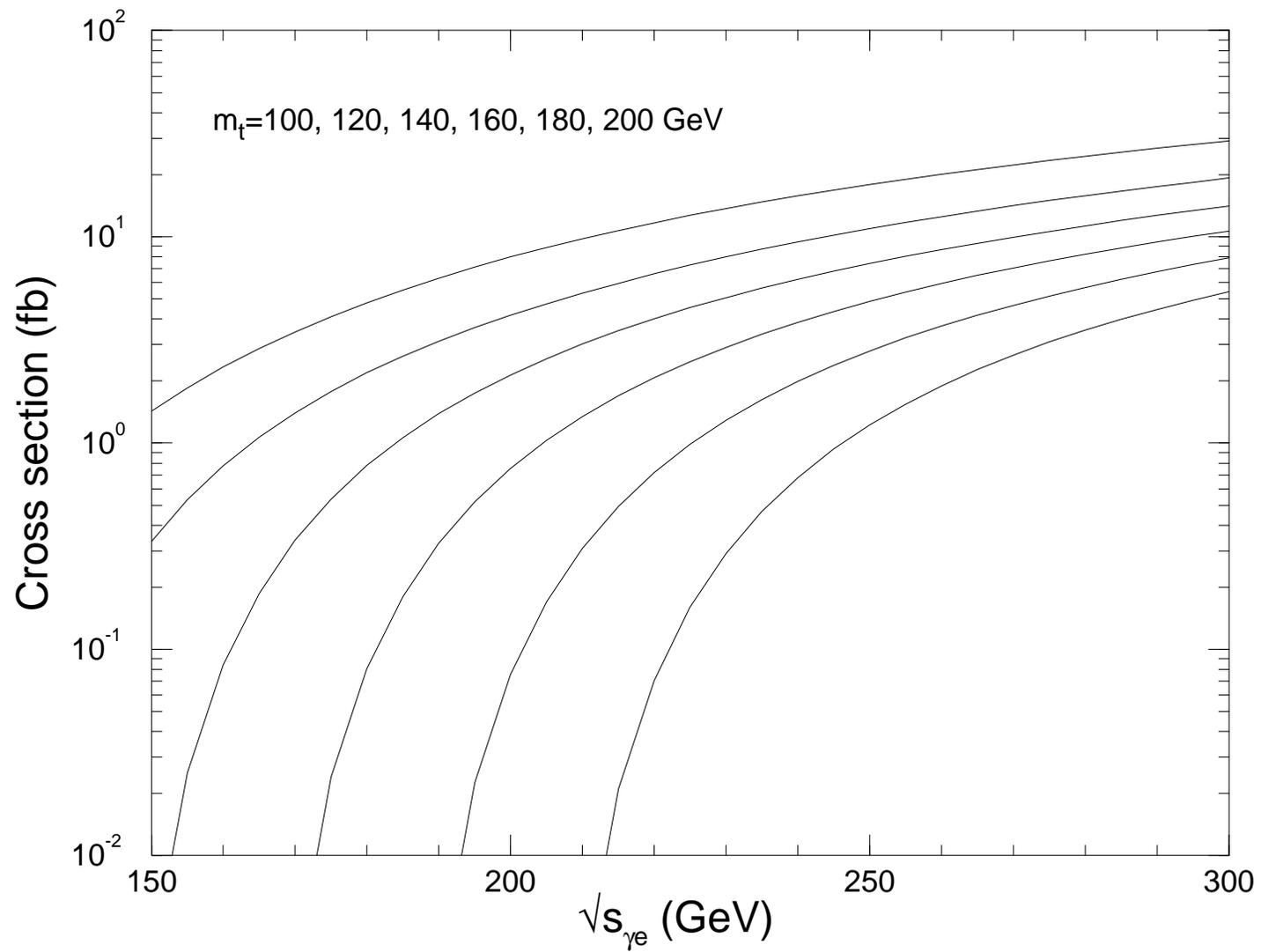
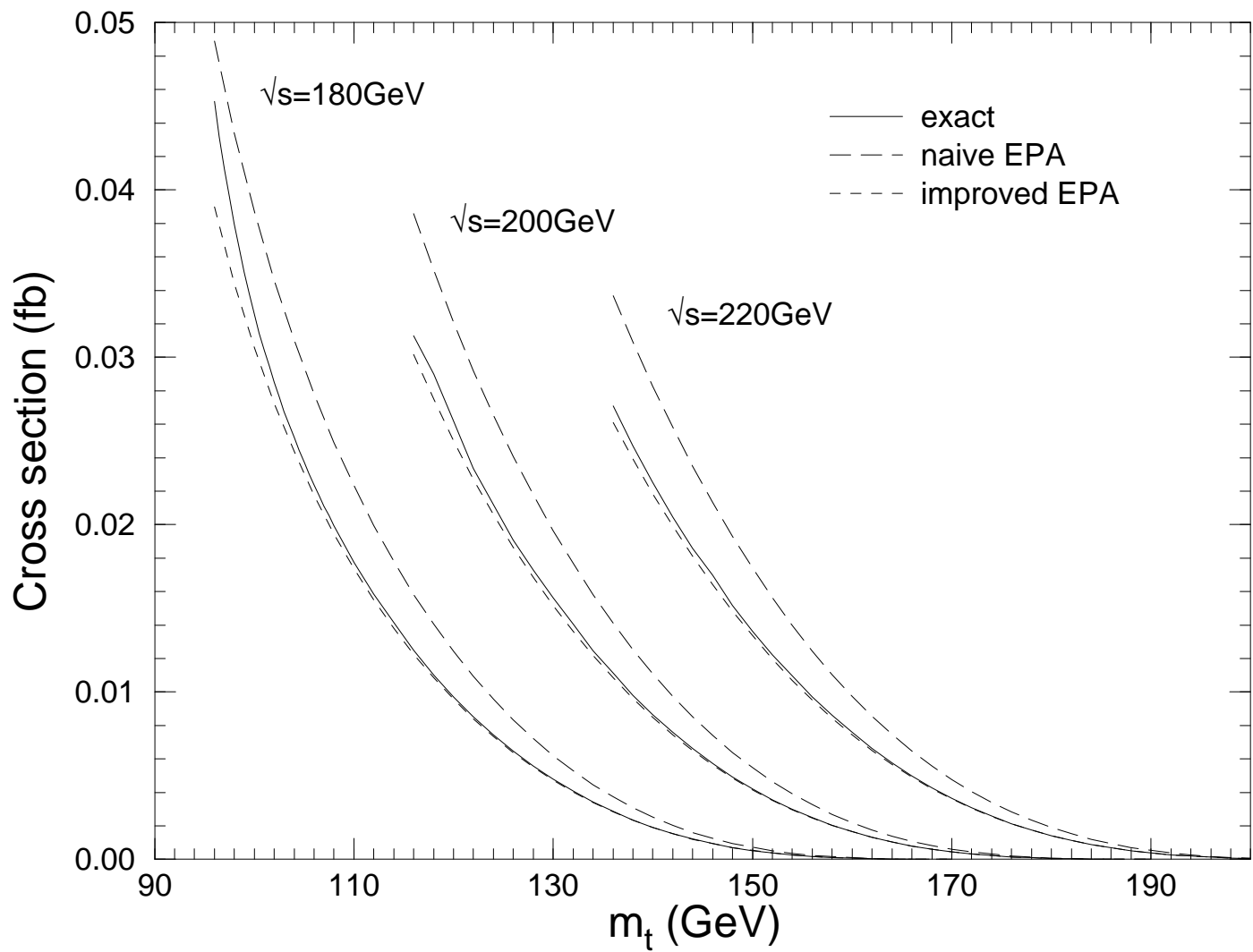
\end{document}